\documentclass[a4paper, 11.5pt]{article}\usepackage[]{graphicx}\usepackage[]{color}
\makeatletter
\def\maxwidth{ %
  \ifdim\Gin@nat@width>\linewidth
    \linewidth
  \else
    \Gin@nat@width
  \fi
}
\makeatother

\definecolor{fgcolor}{rgb}{0.345, 0.345, 0.345}

\usepackage{framed}
\makeatletter
 {\par\unskip\endMakeFramed%
 \at@end@of@kframe}
\makeatother

\definecolor{shadecolor}{rgb}{.97, .97, .97}
\definecolor{messagecolor}{rgb}{0, 0, 0}
\definecolor{warningcolor}{rgb}{1, 0, 1}
\definecolor{errorcolor}{rgb}{1, 0, 0}
\newenvironment{knitrout}{}{} 

\usepackage{alltt}
\usepackage{amsmath, amssymb}
\usepackage{mathtools}
\usepackage[toc, page]{appendix}
\usepackage[english]{babel}
\usepackage{booktabs} 
\usepackage{doi}
\usepackage{longtable}
\usepackage{multirow}
\usepackage{enumitem} 
\usepackage[T1]{fontenc}
\usepackage{graphics}
\usepackage{helvet}
\usepackage[utf8]{inputenc}
\usepackage{lscape}
\usepackage{nameref}
\usepackage[super, comma, sort&compress]{natbib}

\usepackage{tabu} 
\usepackage{textcomp}
\usepackage{todonotes}
\usepackage[sc]{mathpazo}
\usepackage{rotating} 
\usepackage{setspace}
\usepackage{systeme}
\usepackage{xparse}   
\usepackage{xcolor}
\usepackage{booktabs}
\newcommand{\hl}{\textcolor{black}}
\newcommand{\latin}[1]{\textit{#1}}

\newcommand{\abk}[1]{\mbox{#1}\xdot}
\DeclareRobustCommand\xdot{\futurelet\token\Xdot}
\def\Xdot{\ifx\token\bgroup.\else\ifx\token\egroup.\else
  \ifx\token\/.\else\ifx\token\ .\else\ifx\token!.\else
  \ifx\token,.\else\ifx\token:.\else\ifx\token;.\else
  \ifx\token?.\else\ifx\token/.\else\ifx\token'.\else
  \ifx\token).\else\ifx\token-.\else\ifx\token+.\else
  \ifx\token~.\else
  \ifx\token.\else.\ \fi\fi\fi\fi\fi\fi\fi\fi\fi\fi\fi\fi\fi\fi\fi\fi}
\newcommand{\ie}{\abk{\latin{i.\,e}}}

\definecolor{RoyalPurple}{RGB}{106, 26, 74}
\definecolor{MidnightBlue}{RGB}{0,0,153}
\definecolor{BrickRed}{RGB}{153,0,0}

\DeclareMathOperator{\Nor}{N} 
\DeclareMathOperator{\TN}{TN} 
\DeclareMathOperator{\se}{se}

\newcommand{\hlrev}{\textcolor{black}}

\usepackage{geometry}
\usepackage{hyperref}  
\hypersetup{
  bookmarksopen=true, 
  breaklinks=true,
  pdftitle={Combining Evidence from Clinical Trials in Conditional or 
Accelerated Approval}, 
  pdfauthor={Deforth, Micheloud, Roes, Held},
  pdfsubject={},
  pdfkeywords={},
  colorlinks=true,
  linkcolor=RoyalPurple,
  anchorcolor=black,
  citecolor=MidnightBlue,
  urlcolor=BrickRed,
}
 
\title{{Combining Evidence from Clinical Trials in Conditional or 
Accelerated Approval}}
\author{\large \textbf{Manja Deforth$^{1*}$, Charlotte Micheloud$^{1*}$,
    Kit C Roes$^2$, Leonhard Held$^1$}\\ \\
\normalsize$^1$Department of Biostatistics at the Epidemiology, Biostatistics 
and Prevention Institute (EBPI) \\ \normalsize and Center for Reproducible 
Science (CRS), University of Zurich, Zurich, Switzerland \\
\normalsize $^2$Department of Health Evidence, Section Biostatistics, 
Radboud University Medical Center,\\
\normalsize Radboud University, The Netherlands \\ 
\normalsize$^*$Manja Deforth and Charlotte Micheloud contributed equally to this work.    \\\\
\normalsize \textbf{Correspondence}\\ \normalsize Manja Deforth, University of 
Zurich, Department of Biostatistics at the Epidemiology, \\
\normalsize Biostatistics and Prevention Institute (EBPI), Hirschengraben 84, 
8001 Zurich, Switzerland \\ 
\normalsize Email: \href{mailto:manjaelisabeth.deforth@uzh.ch}
{manjaelisabeth.deforth@uzh.ch} \\ \\
\normalsize \textbf{Funding information}\\
{\normalsize This work was \hl{partially} funded by the Swiss National Science
Foundation (project 189295).} \\ \\
\normalsize \textbf{Software and data availability} \\
{\normalsize All analyses were performed in the R programming language version 
4.1.2 (2021-11-01).} \\
{\normalsize Data and code to reproduce all analyses are available at} \\
{\normalsize \href{https://gitlab.uzh.ch/charlotte.micheloud/repro-material-conditional-drug-approval}
{https://gitlab.uzh.ch/charlotte.micheloud/repro-material-conditional-drug-approval}}}  
\date{\today}

\usepackage{fancyhdr}
\usepackage{ulem}
\IfFileExists{upquote.sty}{\usepackage{upquote}}{}
\begin{document}

\onehalfspacing

\maketitle



\begin{center}
  \begin{minipage}{16cm}
  \rule{\textwidth}{0.5pt} \\
  {\textbf{Abstract} \\ 
    Conditional (\hlrev{European~Medicines~Agency)}  or accelerated (\hlrev{U.S. Food and Drug Administration}) approval \hlrev{of drugs} 
    allow earlier access \hlrev{to promising new treatments that address unmet medical needs.}
    Certain \hlrev{post-marketing} requirements must 
    \hlrev{typically be met in order to obtain full approval, such as conducting a new post-market clinical trial.}
 We study the applicability of the recently developed harmonic mean $\chi^2$-test 
 to this conditional \hlrev{or accelerated}
 approval framework. The proposed approach can be used both \hlrev{to support} the design of the post-market trial and the analysis of 
 the combined evidence provided by both trials. Other methods considered are the
 two-trials rule, Fisher's criterion and Stouffer's method. In contrast to some of the traditional methods, the harmonic mean 
$\chi^2$-test always requires a post-market clinical trial. If the $p$-value from the pre-market clinical trial is $\ll 0.025$,
a smaller sample size for the post-market clinical trial is needed than with 
the two-trials rule. For illustration,
 we apply the harmonic mean $\chi^2$-test to a drug which received conditional 
 (and \hlrev{later full}) market licensing by the EMA. A simulation study is 
 conducted to study the operating characteristics \hlrev{of the harmonic mean $\chi^2$-test and two-trials rule} in more detail. We finally 
 investigate the applicability of \hlrev{these two methods} to compute the power at interim of an ongoing post-market
trial. \hlrev{These results are expected to aid in the design and 
assessment of the required post-market studies in terms of the level 
of evidence required for \hlrev{full approval}.}

  }

\rule{\textwidth}{0.5pt} \\
\\
\textbf{KEYWORDS}: \\
\hlrev{Accelerated approval}, conditional approval, harmonic mean chi-squared test, two-trials rule, 
interim power, sample size calculation
\end{minipage}
\end{center}

\newpage

\section{INTRODUCTION}
\label{sec:introduction}
\hlrev{There are several ways how a new treatment can obtain market 
approval by regulatory agencies. \hlrev{ Confirmatory clinical evidence for 
the effectiveness of a drug 
is usually required, and often includes some evidence of robustness.}
Such robustness
  can be demonstrated by direct replication (\hlrev{applying} the two-trials 
  \hlrev{rule})
  or replication in the form of more than one pivotal trial that may be 
  targeted at different endpoints. 
  \hlrev{W}hen a disease is life-threatening or severely debilitating with an unmet 
medical need, or when it is rare,
the European~Medicines~Agency~(EMA) may 
grant a conditional approval.\citep{EMA2016Guideline} 
\hlrev{Likewise, the U.S.~Food and Drug Administration (FDA)
has installed processes and conditions that can allow 
accelerated approval.\cite{FDAGuidance}} This will
give patients earlier access to promising new treatments\hlrev{.} \hlrev{A conditional or accelerated \hlrev{approval, however},} typically requires
that the evidence for a treatment effect from a pre-market clinical trial is 
substantiated in a post-approval clinical study, that may focus
on more relevant clinical endpoints, such as mortality}. 

An example is the drug Fampridine, a drug for patients 
with Multiple Sclerosis. Fampridine received conditional approval in 
July 2011 and \hlrev{full} \hlrev{approval} in May 2017.\cite{EMAFampyra} 
The EMA requirement for \hlrev{full} \hlrev{approval} was ``To conduct a 
double-blinded, placebo-controlled, long-term efficacy and safety study to 
investigate a broader primary endpoint clinically meaningful in terms of 
walking ability and to further evaluate the early identification of responders 
in order to guide further treatment [\dots]".\cite{Fampyra:AssessReport}$^
{\text{(p.~86)}}$ The results from the pre- and post-market clinical trials 
are given in Table \ref{Tab:Fampyra}.
The primary outcome was improvement in walking speed (T25FW
Test) in the pre-market clinical trial and improvement in walking
ability (MSWS-12 score) in the post-market clinical trial.
\cite{Fampyra:AssessReport, Hobart2019} \hl{Such a change in the primary
  outcome between the pre- and post-market clinical trial is quite
  common in the conditional drug approval framework. Specifically, a
  surrogate outcome is often used in the pre-market clinical trial, whereas
  the follow-up time is
  usually longer and the evaluation of a clinical outcome 
  is preferred in the post-market clinical trial.}

\small
\begin{table}[!htp]
\caption{Results from pre- and post-market clinical trials on the efficacy of 
Fampridine for the treatment of patients suffering from Multiple Sclerosis.}
\label{Tab:Fampyra}
\begin{tabular}{p{2.5cm}p{1cm}p{2.8cm}p{4.5cm}p{3cm}} 
\hline
\multirow{2}{*}{\textbf{Clinical trial}} &
\multirow{2}{*}{\textbf{Time}} &
\multirow{2}{*}{\textbf{No. of Patients}} &
\multirow{2}{*}{\textbf{Primary Outcome}} &
\multirow{2}{*}{\textbf{Estimates (in \%)}} 
\\

&
&
&
&
\\ \hline

pooled analysis \newline (MS-F202/3/4) & 
pre & 
Fampridine: 394 \newline Placebo: 237 \newline Total: 631 & 
difference in responder rates (improved walking speed) & 
Fampridine: 37.3   \newline
Placebo: 8.9      \newline                              
Difference: 28.4 \newline
95\% CI: 22.1 to  34.2                                
\\ \hline

218MS305 \newline phase III & 
post & 
Fampridine: 315 \newline Placebo: 318 \newline Total: 633 & 
difference in responder rates (improved walking ability) & 
Fampridine: 43.2   \newline
Placebo: 33.6      \newline                              
Difference: 9.5 \newline
95\% CI: 1.9 to  17               
\\ \hline

\multicolumn{5}{p{15cm}}{Source: \cite{Fampyra:AssessReport, Hobart2019}. 
Shown are unadjusted risk differences with 95\% CIs \cite{Altman.etal2000}. } \\
\end{tabular}
\end{table}
\normalsize


From a regulatory point of view, the question may arise if a
conditional \hlrev{approval} can be granted \hlrev{by the EMA} and, during reevaluation, if the initial
conditional licensing decision was correct. \hlrev{Likewise, in the \hlrev{FDA}
  setting of accelerated approval, the question is whether the initial
  approval can continue or needs to be modified or revoked.}   
  \hlrev{Furthermore, in the design phase of the required post-market trial 
  it is of interest to know the level of evidence which is required for full 
  approval.} 

  Here we consider the \hlrev{most
  straightforward} scenario that pre- and post-market trials
\hlrev{are independent randomized controlled trials.}
In this case, there are several ways to quantify the
overall evidence provided by the two trials. Traditional methods
include for example the two-trials rule, Fisher's criterion, and
Stouffer's method.  \hlrev{Our first objective is to study} the applicability 
of the recently developed harmonic mean $\chi^2$-test\citep{Held2019} to 
\hlrev{assess the statistical level of evidence associated with the totality of 
clinical trials performed up to full approval}
and to
compare it with the more traditional methods mentioned above 
(see Section~\ref{chap:StatisticalMethods}). \hlrev{A comparison and application of the different methods to the Fampridine results from Table
\ref{Tab:Fampyra} illustrates that both Fisher's criterion and Stouffer's method
are not suitable in the conditional drug approval setting, as they may
not require to conduct a post-market trial at all if the pre-market
trial was already very convincing. This is not the case for 
the harmonic mean $\chi^2$-test and the
two-trials rule. In Section~\ref{chap:SimulationStudy}, we therefore study the performance of these two methods 
in an extensive simulation study.}  The harmonic mean $\chi^2$-test and the
two-trials rule may not only be used
\hlrev{for assessing the level of evidence across} the two clinical
trials \hlrev{once completed}, but also at a post-market interim analysis \hlrev{to
  possibly stop} the post-market clinical trial \hlrev{for futility. Our second objective is therefore to investigate the applicability of the two methods for post-market interim analysis} \hlrev{(}see
Section~\ref{PowerAtInterim}\hlrev{)}.  \hlrev{The conclusions are discussed} in
Section~\ref{chap:Discussion}.

\section{STATISTICAL METHODS}
\label{chap:StatisticalMethods}
A short introduction to some notation and assumptions is given in Section 
\ref{MethodNotationAssumptions}. The 
statistical methods, namely the
two-trials rule 
(Section \ref{TwoTrialsRule}), the harmonic mean $\chi^2$-test 
(Section \ref{MethodHarmonicMean}),  Fisher's criterion 
(Section~\ref{Fisher's criterion}) and Stouffer's method 
(Section \ref{Stouffer}) are summarized and compared (Section \ref{MethodsComparison}).

\subsection{Notation and assumptions}
\label{MethodNotationAssumptions}
\hlrev{We focus on balanced two-armed RCTs dealing with life-threatening and 
severely debilitating diseases, and do not consider 
rare conditions, where sample sizes may be very small. Therefore, we} assume a normal distribution with
known variance for the estimate $\hat\theta$ of the true treatment effect $\theta$:
$\hat{\theta} \sim \Nor(\theta, \se^2)$. Here, $\se=\sqrt{2} \sigma/\sqrt{n}$
denotes the standard error of $\hat{\theta}$ based on a sample of size 
$n$ (per group) and a residual standard deviation $\sigma$ (assumed to be the 
same in both groups).
It applies to binary outcomes (where $\theta$ is the log odds ratio), 
survival outcomes (where
$\theta$ is the log hazard ratio) and continuous outcomes (where $\theta$ 
is the mean difference). The exact form of $\sigma$ and $n$ for the different 
types of outcomes can be found in Spiegelhalter, Abrams and Myles
\cite{Spiegelhalter2004} in Section 2.4.

We consider a pre-market clinical trial (index $1$) and an
independent post-market
clinical trial (index $2$) with two groups each to estimate the corresponding
true, unknown treatment effects $\theta_1$ and $\theta_2$, respectively.
\hl{The pre- and post-market clinical trials might have different primary 
outcomes (for example a surrogate and a clinical outcome) and therefore 
different treatment effects, \ie $\theta_1 \neq \theta_2$.
Another possibility is that the primary outcome 
is identical, but the treatment effect $\theta_2$ is different in the post-market 
clinical trial due to a different 
study population.  \hlrev{We assume that both outcomes $\theta_1$
and $\theta_2$ are clinically relevant for the disease and have been selected with care.} }  

In this paragraph, let $i \in \{1, 2\}$ represent the index of the pre- and post-market clinical trials, respectively.
The estimated treatment effect of each trial is
denoted by $\hat{\theta}_{i}$ and is normally distributed with known variance: 
$\hat{\theta}_i~\sim~\Nor(\theta_i, \se^{2}_i)$. The corresponding standard error 
is $\se_i = \sqrt{2}\sigma_i/\sqrt{n_i}$, with the corresponding 
sample size per group $n_i$. The variance ratio of the pre- to post-market trial
$c$ is then equal to $\se^{2}_1 / \se^{2}_2$.  
The test statistic of each trial is 
$z_{i} = \hat{\theta}_{i}/{{\se}_{i}}$, and 
the one-sided $p$-values for the null hypothesis $H_0$: 
$\theta_i = 0$ vs.~the alternative $H_1$: $\theta_i > 0$
is calculated as $p_i=1-\Phi(z_i)$,
where $\Phi(\cdot)$ denotes the cumulative distribution function of 
the standard normal distribution. 
\hlrev{As long as the alternative hypotheses are oriented in the same direction 
in both trials and the point null hypothesis is the same, 
the $p$-value combination methods discussed below, namely the 
two-trials rule, the harmonic mean $\chi^2$-test,  Fisher's criterion, and Stouffer's method, can be used.
Importantly, the method must be selected before the first $p$-value 
has been observed.}



%
\subsection{Two-trials rule}
\label{TwoTrialsRule}
\hlrev{The two-trials rule comes from the FDA requirement to have ``at least two adequate and
well-controlled studies, each convincing on its own, to establish
effectiveness''.\citep{FDA1998}$^{\text{(p.~3)}}$  
While this is not an explicit requirement from the EMA, a Points to Consider 
document \citep{PtC}$^{\text{(p.~5)}}$  makes it implicitly clear that usually
more than one pivotal trial is expected as confirmatory proof of efficacy 
for approval.} 

In the conditional
drug approval framework the two-trials rule can be 
fulfilled by independently
replicating the result of a pre-market clinical trial in a post-market clinical trial,
both significant at the one-sided level $\alpha=0.025$.  The overall
Type-I error rate of this simple decision rule is hence
$\gamma=\alpha^2=0.000625$.

The sample size of the post-market clinical trial can be calculated such
that the power to detect $\hat\theta_1$ at level $\alpha$ reaches a
pre-specified value $1-\beta$. 
In practice, the estimate $\hat\theta_1$ is often inflated 
as compared to the true $\theta_1$ due to various phenomena such as
publication bias and the
winner's curse.\cite{ioannidis2008, Rothwell2021} The latter
will be explained in more detail in Section~\ref{chap:SimulationStudy}. 
\hl{As a result, calculating the sample size of the post-market trial to detect 
$\hat\theta_1$ might lead to an underpowered trial.
A simple method to address this has been proposed in the context of 
replication studies.\cite{MicheloudHeld2022}  In short, 
a shrinkage factor $s$ can be chosen and represents the 
scepticism about the estimate $\hat\theta_1$ from the 
pre-market trial, with $s = 0$ representing no expected
inflation of $\hat\theta_1$.
It follows that the required variance ratio $c$ 
(pre- to post-market squared standard errors 
$\se_1^2/\se_2^2$) only depends on the $z$-value
$z_1$ from the pre-market clinical trial \cite{MicheloudHeld2022} and 
the shrinkage factor $s$:}
\begin{equation}\label{eq:sampleSize}
  c = \Phi^{-1}\left((z_{1-\beta} + z_{1-\alpha})^2/\hl{((1-s)^2z_1^2)}\right),
\end{equation}
\hl{where $z_{1-p}=\Phi^{-1}(1-p)$.}

\hl{The sample size of the post-market trial can then be calculated 
as $n_2 = c n_1 (\sigma_2^2/\sigma_1^2)$, where $n_2$ and $n_1$
are both either samples sizes per group or in total.
If the pre- and post-market standard deviations $\sigma_1$ and $\sigma_2$, 
respectively, 
are the same, the variance ratio $c$ reduces to the \textit{relative} sample size 
$c = n_2/n_1$. 
Formula~\eqref{eq:sampleSize} can also be used when different endpoints
are of interest in the pre- and post-market clinical trials, \ie when
$\theta_1 \neq \theta_2$. To do so, the researchers need to judge 
the expected shrinkage $s$ in the standardized mean difference 
$\hat\theta_2/\sigma_2$ of the post-market 
trial as compared to the standardized
mean difference $\hat\theta_1/\sigma_1$ of the pre-market trial 
and use it in~\eqref{eq:sampleSize}.}
\hlrev{More sophisticated adjustments based on the correlation between 
the two endpoints of interest exist. \cite{Rothwell2021}$^{(\text{Sec.~4.1)}}$
Moreover, 
if the endpoints are very different, it may be preferable to use 
a standard sample size calculation formula based on a 
pre-defined minimal clinically important difference 
\cite{Matthews2006}$^{(\text{Sec.~3.3)}}$
for the post-market sample size. 
}

In the two-trials rule, 
the required significance level $\alpha$ does not depend on the result 
from the pre-market clinical trial. 
\subsection{Harmonic mean $\chi^2$-test}
\label{MethodHarmonicMean}
The harmonic mean $\chi^2$-test was developed to combine the $p$-values from 
independent clinical trials with the same research hypothesis into one 
overall $p$-value.\citep{Held2019} 
The test statistic $X^2$ of the 
harmonic mean $\chi^2$-test
is twice the harmonic mean $t_H^2$ of the squared test-statistics:
\begin{align}
    X^2 &= 2 \, t_H^2 = \frac{4}{1/z_1^2 + 1/z_2^2}. \label{formula:X} 
\end{align}
Weights $w_1$ and $w_2$ can also be introduced, then
\begin{align}
    X^2 & = \frac{w^2}{w_1/z_1^2 + w_2/z_2^2} \text{ with } w = \sqrt{w_1} +
    \sqrt{w_2}. \label{formula:Xw}
\end{align}
Under the null hypothesis $H_0$ where $z_1\sim\Nor(0,1)$ and
$z_2\sim\Nor(0,1)$, both \eqref{formula:X} and \eqref{formula:Xw} are 
$\chi^2$-distributed
with one degree of freedom. This property can be used to derive the
one-sided $p$-value $p_{H}$ of the harmonic mean $\chi^2$-test under the assumption
that the alternative hypothesis of each trial 
is $H_1$: $\theta_i > 0$, with $i \in \{1, 2\}$. 
If $z_1>0$ and $z_2>0$ both hold, then 
$p_{H} = \Pr\{\chi^2_1 \geq X^2\}/4 = 
\left[1 - \Phi(X)\right]/2$ where $X=+\sqrt{X^2}$. Likewise
we can obtain the critical value 
\[
  c_H = \chi^2_1 (1-4 \gamma) 
\]
for the test statistic \eqref{formula:X} and~\eqref{formula:Xw},
where $\gamma$ is the required significance level and
$\chi^2_{\tiny \nu}(.)$ denotes the quantile function of the
$\chi^2$-distribution with $\nu$ degrees of freedom.

The $p$-value from the harmonic $\chi^2$-test is usually compared against 
$\gamma=\alpha^2$, the overall Type-I error rate of the two-trials rule at 
one-sided significance level $\alpha$, with critical value 
$c_H = 9.14$ for $\gamma = 0.025^2$.

The harmonic mean $\chi^2$-test can also be applied to determine the
required $z$-value $z_2$  (respectively $p$-value $p_2$) in the post-market trial
given the $z$-value $z_1$ (respectively $p$-value $p_1$) from the
pre-market clinical trial, to achieve overall significance at level
$\gamma$. For the more general weighted version, rearranging 
\eqref{formula:Xw} yields the requirement
\begin{eqnarray}\label{eq:threshold}
z_2 & \geq & \bar z_2 = \sqrt{w_2}/\sqrt{w^2/c_H - w_1/z_1^2}.
\end{eqnarray}
This implies an upper bound
$\bar p_2 = 1 - \Phi(\bar z_2)$ for the $p$-value $p_2$ from the
post-market trial to achieve overall significance. 
The necessary requirement $p_2 \leq 0.065$ for
overall significance in the unweighted case at level $\gamma = 0.025^2$ can
be obtained from \eqref{eq:threshold} for $w_1=w_2=1$ (so $w=2$) and
$z_1^2 \rightarrow \infty$, where
$\bar z_2 \downarrow \sqrt{9.14/2^2} = 1.51$ and so
$\bar p_2 \uparrow 0.065$.  This means that, no matter how convincing the pre-market clinical trial is, the post-market clinical trial
always needs to be conducted and has to provide sufficient evidence
against the null hypothesis on its own.

\hlrev{How to weigh the trials is a decision that belongs to 
the regulators. We argue that it might be sensible to give 
more weight to the pre-market trial as it is the study which led to 
conditional approval.}
\hlrev{For this reason,} we will also consider the case where the weight of
the pre-market trial
is 60\% rather than 50\%, then the necessary requirement on $p_2$ is 
$p_2 \leq 0.087$. The upper bound $\bar p_2$ can be used as an adaptive significance
level $\alpha=\alpha(p_1)$, which depends on the result from the pre-market trial, in the sample size
computation \eqref{eq:sampleSize} for the post-market trial. The necessary condition on $p_1$ can also be derived from 
\eqref{eq:threshold}. The upper bound on $p_1$ 
is $\bar p_1 = 0.065$ in the unweighted case
and $\bar p_1 = 0.048$ in the weighted case.


\subsection{Fisher's criterion}
\label{Fisher's criterion}
Another way to combine $p$-values is Fisher's criterion.\cite{Fisher-1958, Bauer1994} Overall significance is achieved if
$p_1 p_2 \leq c_F = \exp\{-\frac{1}{2}\chi^2_4(1-\gamma)\}$, where
$\gamma=\alpha^2$. Given the $p$-value $p_1$ from the pre-market
trial, the post-market trial $p$-value $p_2$ has to fulfill
$p_2 \leq \bar p_2 = c_F/p_1$ to achieve overall significance. Similar to the harmonic mean $\chi^2$-test, the sample size calculation \eqref{eq:sampleSize} for 
the post-market trial is now based on the adaptive significance level $\bar p_2$. 

If the evidence from the pre-market clinical trial is large, $p_1$ is
sufficiently small and Fisher's bound $\bar p_2$ will be larger than
one. For example, for $\gamma=0.025^2$ we have
$c_F=0.000058$.  If
the pre-market $p$-value $p_1$ is smaller than $c_F$, the post-market
clinical trial does not need to be conducted at all.  This is an
undesired property for conditional drug approval, where regulators
usually want additional evidence from a post-marketing trial, no
matter how convincing the pre-market trial was.  \hl{In some cases,
  drug approval may even be granted based on one study with
  overwhelming evidence.\cite{Fisher1999, Shun2005} Using only one
  study might be viewed as an application of Fisher's criterion;
  however in this case the replication requirement is dropped.}


\subsection{Stouffer's method}
\label{Stouffer}
Another $p$-value combination approach  is Stouffer's method, also known as 
inverse-normal method, which
is based on the $z$-values $z_i = \Phi^{-1}(1-p_i)$. 
Under the assumption of no effect, the test statistic
$z = (z_1 + z_2) / \sqrt{2}$ follows a standard normal
distribution. Overall significance at level $\gamma$ is achieved if
$z \geq z_{1-\gamma}=\Phi^{-1}(1-\gamma)$. Weights can also be introduced. 


Suppose the $p$-value $p_1$ with corresponding $z$-value $z_1$ from
the pre-market clinical trial is known. The post-market $z$-value $z_2$ then
needs to fulfill $$z_2 \geq \sqrt{2} \, z_{1-\gamma} - z_1$$ to achieve
overall significance at level $\gamma$. This can be translated to an
upper bound for the $p$-value $p_2$ from the post-market trial:
\[
  p_2 \leq \bar p_2 = 1 - \Phi(\sqrt{2} \, z_{1-\gamma} - z_1),
\]
which serves as an adaptive significance level in the sample size
calculation \eqref{eq:sampleSize} for the post-market trial. For 
$\gamma=0.025^2$ we have 
$\bar p_2 = 1-\Phi(4.56-z_1)$, so the upper bound can be close to 1 if the pre-market $z$-value $z_1$ is \hlrev{very} large.

\subsection{Comparison of the different methods}
\label{MethodsComparison}
\begin{figure}
\centering
\begin{knitrout}
\definecolor{shadecolor}{rgb}{0.969, 0.969, 0.969}\color{fgcolor}
\includegraphics[width=\maxwidth]{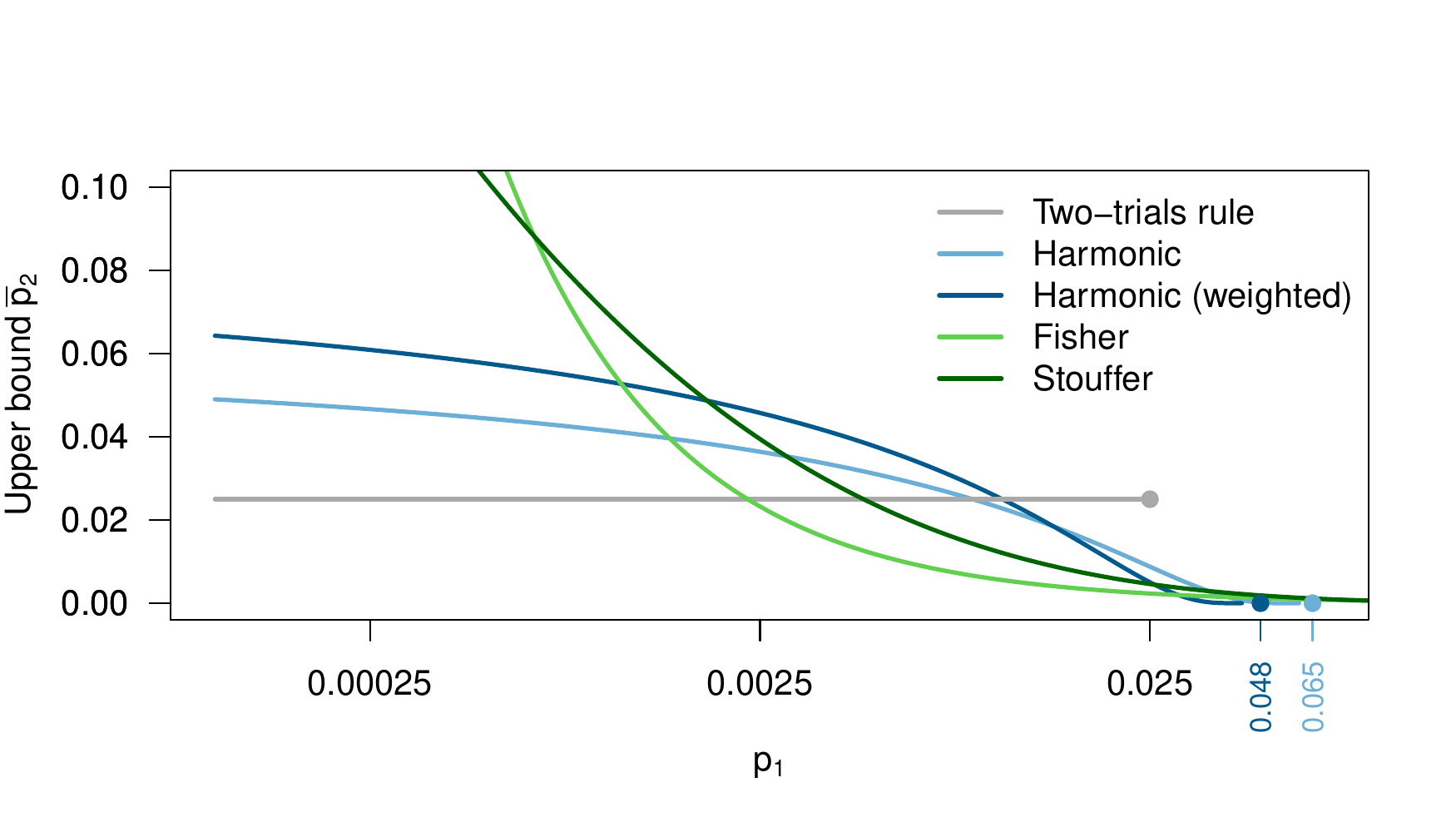} 
\end{knitrout}

\begin{knitrout}
\definecolor{shadecolor}{rgb}{0.969, 0.969, 0.969}\color{fgcolor}
\includegraphics[width=\maxwidth]{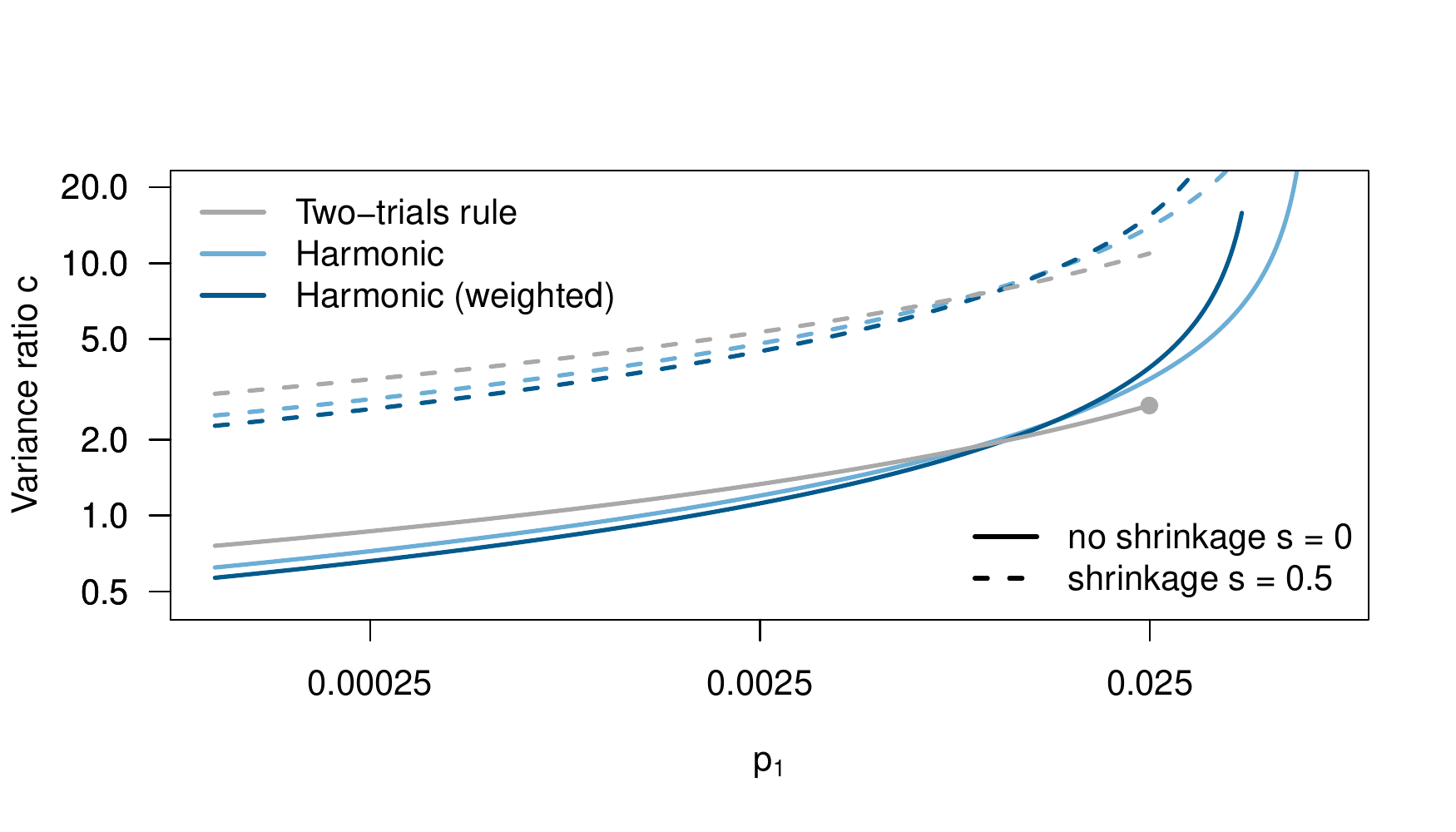} 
\end{knitrout}
\caption{The upper bound
for the 
post-market clinical trial $\bar p_2$ \hlrev{(top)} and the variance ratio $c$ \hlrev{(bottom)} 
calculated with the different methods. 
The upper bound $\bar p_2$ is calculated based on the $p$-value $p_1$
of the pre-market clinical trial. In the weighted version of the harmonic 
mean $\chi^2$-test, the weights are $w_1 = 3$ and $w_2 = 2$.}
\label{FigLevRelSS}
\end{figure}

A comparison of the methods described above is given in Figure \ref{FigLevRelSS}
(upper panel).
For each method, the upper bound post-market $p$-value ($\bar p_2$)
is shown as a function of the $p$-value ($p_1$) 
from the pre-market clinical trial. 
This means that overall significance is achieved if and only if
$p_2 \leq \bar p_2$. We set the significance level to $\alpha = 0.025$.   
If $p_1 \leq 0.025$, the two-trials rule is fulfilled for any 
$p_2 \leq 0.025$, regardless of the value of $p_1$. 
For the other three methods (harmonic mean $\chi^2$-test, Fisher's criterion and
Stouffer's method), the smaller $p_1$, the larger $\bar p_2$. In other 
words, the more convincing the pre-market clinical trial, the less evidence 
is required in the post-market clinical trial. However, as mentioned
earlier, in case of a 
very small $p_1$ there would be no need for a post-market clinical trial with 
Fisher's criterion.
This is an unwanted property in the conditional approval framework where a 
post-market trial is always required to confirm the result of the pre-market trial.
Stouffer's method will always require a post-market trial, 
but the adaptive significance level $\bar p_2$ can become very large. 
Due to this, we decided to not investigate these two methods further. 
The harmonic mean $\chi^2$-test does not suffer from this drawback:
if the necessary condition on $p_1$ is fulfilled  ($p_1 \leq 0.048$ for 
$w_1$ = 3 and $w_2$ = 2, or 
$p_1 \leq 0.065$ in the unweighted case), the harmonic mean $\chi^2$-test
always requires a post-market trial.

\hl{Figure \ref{FigLevRelSS} (lower panel) shows the variance ratio $c$
calculated with~\eqref{eq:sampleSize} for the two-trials rule,
and the 
unweighted and weighted harmonic mean $\chi^2$-test
(with the suitable significance level)
as a function of the $p$-value $p_1$ of the pre-market clinical trial.
The sample size $n_2$ of the post-market trial can be calculated from $c$
via $n_2 = c n_1 (\sigma_2^2/\sigma_1^2)$, where $n_2$ and $n_1$ are both either 
samples sizes per group or in total. 
Two scenarios are considered: In the first one, no inflation of the effect 
estimate $\hat\theta_1$ is expected ($s = 0$), while a 50\% shrinkage 
($s=0.5$) is used in the calculation of $c$ in the second scenario. 
The power is always
fixed at 90\%. 
Compared to the two-trials rule, the harmonic mean $\chi^2$-test requires a 
smaller sample size for the post-market clinical trial if  
$p_1 \ll \alpha = 0.025$ and a larger sample size if $p_1 \approx 0.025$.
Shrinking the pre-market trial effect estimate $\hat\theta_1$ 
results in a larger $c$ for the same level of power. }

\subsection{\hl{Application}}
We now revisit the Fampridine clinical trial results given in Table
\ref{Tab:Fampyra} to illustrate the applicability of the different
methods to the conditional drug approval setting. We apply the
variance-stabilizing arcsine square root transformation
\cite{Matthews2006}$^{(\text{Sec.~3.4)}}$ to the different proportions
in order to achieve approximate normality with a constant standard
deviation of \hl{$\sigma_1=\sigma_2=0.5$}. The corresponding $z$-values (and
$p$-values) of the pre- and post-market clinical trial turn out to be
$z_1=8.6$
($p_1< 0.00001$) and
$z_2=2.5$
($p_2=0.014$). The pre-market $p$-value $p_1$
is so small that Fisher's criterion does not require a post-market clinical trial
at all at the standard two-trials rule significance level
$\gamma=0.025^2$.  Stouffer's method does in principle require a
post-market trial, but the $p$-value threshold
$\bar p_2 = 0.999976$ \hl{is very close to 1}. This illustrates that
both Fisher's and Stouffer's methods are not suitable for the
conditional drug approval setting.
In contrast, application of \eqref{eq:threshold} with 
$z_1=8.6$ and 
$\gamma=\alpha^2=0.025^2$ shows that the
harmonic mean $\chi^2$-test has the 
requirement $p_2 \leq 0.062$
(respectively
$p_2 \leq 0.083$ in the
weighted case) for the post-market $p$-value $p_2$. 
Both requirements are met by the observed post-market $p$-value
$p_2=0.014$.


It is also of interest to compare the sample size calculation based on the
harmonic mean $\chi^2$-test with the traditional two-trials rule.
\hl{The post-market trial has already been conducted, so we are in a position
to reconstruct the post-market sample size calculation based on the almost universal 
$\alpha=0.025$ standard
and to compute the required sample size if the 
adjusted level \hl{$\bar p_2$} based on \eqref{eq:threshold} would had been used instead.}
The post-market clinical trial aimed to reach a sample size of 590
patients in total, allowing for 15\% drop out.\cite{Hobart2019} 
The corresponding sample size of 251 per group (before
drop-out) has a power of 90\% to detect the
\hlrev{standardized} effect size
$d=0.29$ 
at the one-sided significance level
$\alpha=0.025$.
\hl{Sample size calculations based on the harmonic mean $\chi^2$-test} 
are summarized in Table \ref{tab:tab1}.  
The evidence from the pre-market clinical trial is very strong (with
$z$-value = 8.6). The adaptive level based on
the unweighted harmonic mean $\chi^2$-test is
\hl{$\bar p_2=$} {0.062}. If this significance level would
have been used in the post-market trial, only {444}
patients would have to be included in the post-market clinical trial. This
corresponds to a reduction in sample size of
{25\%}. The 3:2 weighted harmonic mean
$\chi^2$-test has the adaptive significance level \hl{$\bar p_2=$} {0.083} so the reduction in sample size
would have been even more pronounced
({32\%} reduction down to
{400} patients).
  
\begin{table}[!htb]
\caption{\label{tab:tab1} Sample size calculation for the post-market 
clinical trial based on the two-trials rule (left) and the harmonic mean 
$\chi^2$-test (right). 
}
\begin{center}
\begin{tabular}{llll}
\hline
& \textbf{Two-trials rule} & \multicolumn{2}{c}{\textbf{Harmonic Mean $\chi^2$-Test}} \\ 
& & unweighted & weighted \\ \hline 
  Required significance level & 0.025 & {0.062} & 
  {0.083} \\
  Required total sample size & 590 & {444} & 
  {400} \\
  Sample size reduction & & {25\%}& 
  {32\%} \\ \hline
\end{tabular}
\end{center}
\end{table}

\section{SIMULATION STUDY}
\label{chap:SimulationStudy} 

\paragraph{\hlrev{Set-up}} A simulation study was conducted in order to obtain a better 
understanding of \hlrev{the two-trials rule and (un)weighted harmonic
mean $\chi^2$-test,} and \hlrev{to investigate} how \hlrev{these methods} compare. 
The details are specified below and the simulation 
has been carried out according to the 
recommendations of Burton et al.\cite{Burton2006} and Morris, 
White and Crowther\cite{Morris2019}. The simulation study was conducted in the 
programming language R.\cite{R} The performance of each method was 
measured based on the rejection percentage,\citep{Morris2019} \ie the proportion
of simulations where overall significance can be declared. Under the null, 
the rejection percentage is the Type-I error rate while it is the power in 
the other scenarios. The number of repetitions $n_{\text{sim}}$ for the 
simulation was calculated as follows:\cite{Morris2019}
\begin{align*} 
n_{\text{sim}} = \frac{\text{E}(\text{Power}) \cdot 
(1-\text{E}(\text{Power}))}{(\text{Monte Carlo se}_{\text{req}})^2} \,.
\end{align*} 
With $n_{\text{sim}}=10,000$ repetitions, 
the Monte Carlo standard error $\se_{\text{req}}$ is 
smaller than $0.5$\% for any value of the power. 

\begin{figure}
\centering
\includegraphics[width=1\textwidth]{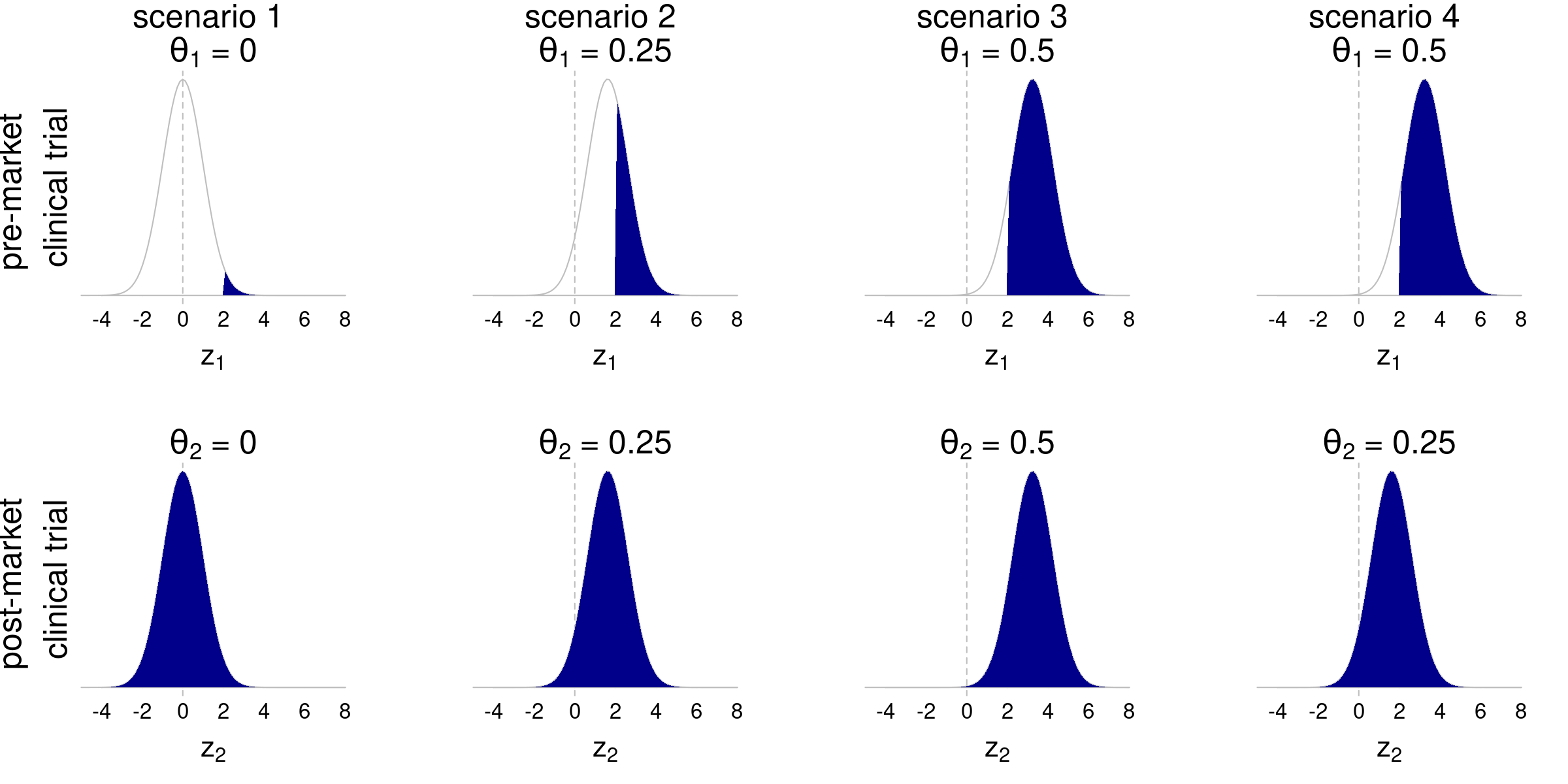}
\caption{Truncated normal distribution of test statistic $z_1$ from pre-market 
clinical data (top) and normal distribution of test statistic $z_2$ from 
post-market clinical data (bottom) at different treatment effects $\theta$ 
(scenario~1 to 4).}
\label{FigDensity}
\end{figure}

One-sided, superiority, balanced two-armed 
(treatment vs. placebo) pre- and post-market clinical trials were simulated 
assuming different values for the true effect $\theta$: $0$, $0.25$ and $0.5$.
We chose four different combinations of $\theta_1$ and $\theta_2$ leading to 
four scenarios (see~Figure \ref{FigDensity}): 1) $\theta_1 = \theta_2 = 0$, 2) 
$\theta_1 = \theta_2 = 0.25$, 3) $\theta_1 = \theta_2 = 0.5$ and 4) 
$\theta_1 = 0.5, \theta_2 = 0.25$. The last scenario represents the case 
where the true effects in the pre-market and the post-market clinical trial are
different, maybe due to a switch from a surrogate endpoint to a clinical outcome. \hl{For simplicity, the standard deviation $\sigma_1 = \sigma_2 
= \sigma = 1$ is assumed to be the same in the pre- and post-market trial, 
and so $c$ can be directly interpreted as the relative sample size $n_2/n_1$.}

Pre-market clinical trials were assumed to have evidence for a treatment effect 
($p_1 \leq \alpha$). 
The test statistics $z_1$ were simulated from a normal distribution
\begin{eqnarray}\label{eq:tn_distribution}
z_1 \sim \TN(\mu, \sigma^2, z_{1-\alpha}, \infty) \, ,
\end{eqnarray}
with mean $\mu = \theta_1 \cdot \sqrt{n_1}/(\sqrt{2} \sigma) = 
z_{1 - \alpha} + z_{1 - \beta}$,\cite{Matthews2006}$^{(\text{Sec.~3.3)}}$
where $1 - \beta$ is the power from the 
pre-market trial, and 
truncated to the interval $[z_{1-\alpha}, \infty)$.

The sample size of $n_1 = 85$ (per group) 
was chosen in such a way that the pre-market 
clinical trial had a power of 90\% to detect 
$\theta_1 = 0.5$ and so the power is reduced to only 
$37$\% for $\theta_1 = 0.25$.

\begin{figure}[htb!]
\centering
\includegraphics[width=1\textwidth]{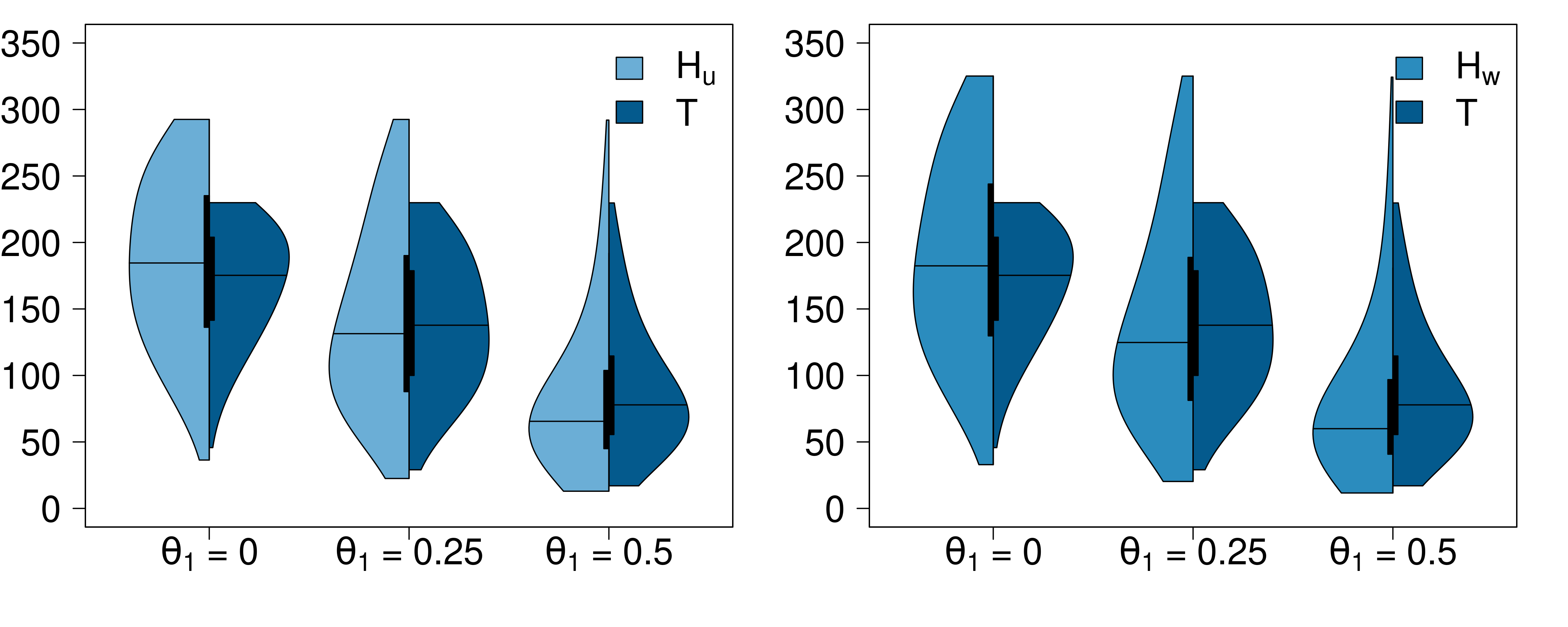}
\caption{Distribution of the sample sizes $n_2$ (per group) of the post-market clinical trial based on 
the estimated treatment effect $\hat\theta_1$ of the pre-market clinical trial and 
90\% conditional power. The different methods are unweighted ($H_u$) 
and weighted ($H_w$) harmonic mean $\chi^2$-test
and the two-trials rule (T).
}
\label{FigSampleSize}
\end{figure}

The sample size of the post-market clinical trial was calculated to detect 
the effect $\hat\theta_1$ estimated in the pre-market clinical trial with a power of 
90\% and with the different methods as discussed in 
Section~\ref{chap:StatisticalMethods}, \hl{assuming no shrinkage ($s = 0$).}
As can be seen in 
Equation~\eqref{eq:sampleSize}, it is equivalent to know the $z$-value $z_1$
instead of the value of the effect estimate $\hat\theta_1$ for 
the sample size calculation.

Figure~\ref{FigSampleSize} compares 
the sample sizes (per group) obtained with the unweighted and
weighted ($w_1 = 3$ and $w_2 = 2$) harmonic mean $\chi^2$-test and the two-trials rule.
For all three methods there is an upper bound for the sample size 
regardless of the value of true effect. It corresponds to draws where the 
one-sided $p$-value $p_1$ from the pre-market clinical trial was borderline 
significant, 
\ie $p_1 \approx 0.025$. The bound based on the weighted harmonic mean $\chi^2$-test 
($n_2^{\mbox{\tiny max}} = 
326$)
is larger than with the unweighted harmonic mean 
$\chi^2$-test ($n_2^{\mbox{\tiny max}} = 293$),
 which is larger compared to 
the two-trials rule ($n_2^{\mbox{\tiny max}} = 
230$). 
With all methods, the median sample size $n_2$ of the post-market clinical trial 
decreases with increasing true effect $\theta_1$. 
The larger the true effect $\theta_1$, the fewer patients are necessary to
detect it. The test statistic $z_2$ of the post-market 
clinical trials was simulated from a normal distribution:  
\begin{align*} 
z_2 \sim \Nor\bigg(\theta_2 \cdot \frac{\sqrt{n_2}}{\sqrt{2} \sigma}, 1\bigg), 
\end{align*} 
with $\theta_2$ and $n_2$ being the treatment effect, and the sample size 
(per group) of 
the post-market clinical trial, respectively. The choice of selecting only 
the significant clinical trials for the pre-market trials and all clinical trials for the 
post-market trials reflects what happens 
in practice: only the persuasive clinical trials are granted conditional approval, 
while the results of all the post-market clinical trials need to be reported.


\paragraph{Type-I error and power of each method}
Table~\ref{tbl:results} displays the rejection rate, which is the 
percentage of study pairs where overall significance is declared,
for each method and under 
each scenario. The median sample sizes (absolute $n_2$ and relative $c$) are 
included for completeness; a more exhaustive picture can be found in 
Figure~\ref{FigSampleSize}.

In the first scenario, the true effect is null in both the pre- and post-market 
clinical trials. The rejection rate hence represents the proportion of 
clinical trials 
where overall significance is, in this scenario, incorrectly declared.
As the pre-market clinical trial has 
$p_1 \leq \alpha$ by construction, the rejection rate under this scenario is not
the overall Type-I error rate, but the Type-I error rate of the post-market trial 
only. 
By definition, the Type-I error rate is $\alpha = 0.025$ with the 
two-trials rule. With the harmonic mean $\chi^2$-test, in contrast, 
the Type-I error rate is smaller than $\alpha$ as only $p_1 \leq \alpha$
were considered. In the unweighted (resp. weighted) case, 
$p_1 \leq 0.065$ (resp. $0.048$) might have led to overall significance.

For the other three scenarios, the same logic prevails: the rejection 
rate can be interpreted as the power of the post-market trial 
(and not the overall power) as the power of the pre-market clinical trial 
is fixed (90\% for scenarios 3 and 4 and 37\% for scenario 2). 
Under every scenarios, the power is the smallest with 
the two-trials rule as compared to the harmonic mean 
$\chi^2$-test. 
One might wonder why the power is always below 90\%, although
every post-market trial was powered for this value. 
This is due to the truncation $p_1 \leq 0.025$, which causes 
a phenomenon called `the winner's curse',\cite{button2013}$^{\text{(p.~366)}}$
which 
``occurs when thresholds, such as statistical significance, are used to 
determine the presence of an effect and is most severe when thresholds are 
stringent and clinical trials are too small and thus have low power''. This 
phenomenon creates estimates $\hat\theta_1$ that are on average overestimated; 
\ie larger than the true effect $\theta_1$. 
\hl{A similar problem occurs when the aim is to conduct a replication of
 a promising sub-group finding in an overall non-significant trial.\cite{Tanniou2016}}
As the post-market
trials were powered to detect $\hat\theta_1$, the sample size $n_2$
is too small and the post-market trials are underpowered.
The effect of the winner's curse on the power is only moderate under 
the third scenario, it is in contrast substantial under the 
second scenario as the pre-market trial is underpowered (37\%).
In addition to this, the fourth scenario suffers
from another problem: the true effects $\theta_1 = 0.5$ and 
$\theta_2 = 0.25$ are not the same, and calculation 
of the sample size $n_2$ did not take this into account, 
resulting in a dramatically low power.

\begin{table}[!h]

\caption{\label{tab:simple-tables}\label{tbl:results} Median sample size $n_2$
            (per group) of
            the post-market clinical trial, median relative sample size $c$, 
            and rejection rate under the four scenarios with the 
            three methods: two-trials rule ($T$), weighted ($H_w$) and
            unweighted ($H_u$) harmonic mean $\chi^2$-test.}
\centering
\begin{tabular}[t]{lrrr}
\toprule
Method & $n_2$ & c & Rejection rate [\%]\\
\midrule
\addlinespace[0.3em]
\multicolumn{4}{l}{scenario 1}\\
\hspace{1em}$H_u$ & 185 & 2.2 & 2.0\\
\hspace{1em}$H_w$ & 182 & 2.2 & 2.2\\
\hspace{1em}T & 175 & 2.1 & 2.5\\
\addlinespace[0.3em]
\multicolumn{4}{l}{scenario 2}\\
\hspace{1em}$H_u$ & 131 & 1.6 & 55.1\\
\hspace{1em}$H_w$ & 125 & 1.5 & 56.2\\
\hspace{1em}T & 138 & 1.6 & 53.7\\
\addlinespace[0.3em]
\multicolumn{4}{l}{scenario 3}\\
\hspace{1em}$H_u$ & 66 & 0.8 & 85.9\\
\hspace{1em}$H_w$ & 60 & 0.7 & 86.2\\
\hspace{1em}T & 78 & 0.9 & 84.8\\
\addlinespace[0.3em]
\multicolumn{4}{l}{scenario 4}\\
\hspace{1em}$H_u$ & 66 & 0.8 & 43.0\\
\hspace{1em}$H_w$ & 60 & 0.7 & 45.2\\
\hspace{1em}T & 78 & 0.9 & 38.3\\
\bottomrule
\end{tabular}
\end{table}

\paragraph{Overall performance}

\hl{As the outcome of the pre-market trial is random, the sample size 
$n_2$ and the power of the post-market trial also have a distribution.
Table~\ref{tbl:results} provides the (median) sample size $n_2$
and (mean) power generated by the two methods; but does not
provide information on the joint distribution of sample size and power.}

In the following we compare the probability that the unweighted
harmonic mean $\chi^2$-test method leads to a smaller sample size
$n_2^{\mbox{\tiny Hu}}$ and simultaneously a larger power
\begin{eqnarray}\label{eq:pow_Hm}
\mbox{power}_{\mbox{\tiny $H_u$}} =
      1 - \Phi\left(\bar z_2 - \frac{\theta \sqrt{n_2^{\mbox{\tiny $H_u$}}}}
      {\sigma \sqrt{2}} \right) \,
\end{eqnarray}
to detect the true effect $\theta$
than the two-trials rule. \hl{For simplicity, we assume
that the outcomes and standard deviations are the same 
in pre- and post-market trials, \ie $\theta_1 = \theta_2 = \theta$ and
 $\sigma_1 = \sigma_2 = \sigma$, and apply no shrinkage 
 in the sample size calculation~\eqref{eq:sampleSize}.}
We denote this probability $\Pr(\mbox{$H_u$ superior})$,
namely the probability that the harmonic mean $\chi^2$-test is superior
to the two-trials rule.
Similarly,  the harmonic mean $\chi^2$-test is inferior
if the sample size $n_2^{\mbox{\tiny 2TR}}$ is smaller and the power
\begin{eqnarray}\label{eq:pow_2TR}
\mbox{power}_{\mbox{\tiny 2TR}} =
      1 - \Phi\left(z_{1 - \alpha} - \frac{\theta \sqrt{n_2^{\mbox{\tiny 2TR}}}}
      {\sigma \sqrt{2}} \right) \,
\end{eqnarray}
to detect the true effect $\theta$ is larger with the two-trials
rule than with the
harmonic mean $\chi^2$-test method.
We also calculate the probability of inconclusive results, \ie 
the harmonic mean $\chi^2$-test method leads to a smaller sample size, 
but also to a smaller power, and the other way around. 

In the calculation of the sample size $n_2$ (or equivalently the relative
sample size $c$),
the only difference between the two approaches is the threshold
for the $z$-value of the post-market trial: $z_{1 - \alpha}$ for the two-trials
rule and $\bar z_2$ for the harmonic mean $\chi^2$-test method
(see~\eqref{eq:threshold}). Using~\eqref{eq:sampleSize}, a smaller sample size
$n_2^{\mbox{\tiny Hu}}$ is required with the harmonic mean $\chi^2$-test
method as compared to the two-trials rule if $\bar z_2 < z_{1- \alpha}$
or equivalently $\bar p_2 > \alpha$.
From~\eqref{eq:threshold}, we see that
$\bar p_2 > \alpha$ holds
whenever $z_1 > 1/\sqrt{4/c_H - 1/z_{1 - \alpha}^2} \coloneqq b$.
If $\alpha = 0.025$, this happens when
$p_1 < 0.009$, as can also be
observed in Figure~\ref{FigLevRelSS}.

Furthermore,  solving
$\mbox{power}_{\mbox{\tiny Hu}} > \mbox{power}_{\mbox{\tiny 2TR}}$
with $\bar p_2 > \alpha$ indicates that
the power of the
post-market trial to detect the \textit{true} treatment effect
is larger with the harmonic mean $\chi^2$-test method
if $z_1 > \mu$, the mean of the normal distribution defined
in~\eqref{eq:tn_distribution}.
The condition for the harmonic mean $\chi^2$-test
to be superior (simultaneously smaller sample size and larger power)
than the two-trials rule is therefore
\begin{equation}\label{eq:HmBetter}
z_1 > \mbox{max}(\mu, b) \, .
\end{equation}
The harmonic mean $\chi^2$-test is inferior to the two-trials rule if
\begin{equation}\label{eq:TTRBetter}
\mu < z_1 < b \, .
\end{equation}

The probabilities of~\eqref{eq:HmBetter}
and \eqref{eq:TTRBetter} only depend on the power $1 - \beta$
from the pre-market clinical trial
(and the level $\alpha$) and can be computed
with numerical integrations as follows:

\begin{eqnarray}
\Pr(\mbox{$H_u$ superior}) & = & \int_{\scriptsize{\mbox{max($\mu$, $b$)}}}^
{\scriptsize{\infty}}
\TN(z_1; \mu, 1, z_{1-\alpha}, \infty) d z_1
\label{eq:hmbetter}\\
\Pr(\mbox{$H_u$ inferior}) &  = & \label{eq:2trbetter}
\begin{cases}
\int_{\scriptsize{\mbox{$\mu$}}}^
{\scriptsize{\mbox{$b$}}} \TN(z_1; \mu, 1, z_{1-\alpha}, \infty)
d z_1,& \mbox{ if $\mu$ < $b$}
\\
0, & \mbox{otherwise},
\end{cases}
\end{eqnarray}
where $\TN$ is the truncated normal distribution defined
in~\eqref{eq:tn_distribution}.

\begin{figure}[!h]
\centering
\begin{knitrout}
\definecolor{shadecolor}{rgb}{0.969, 0.969, 0.969}\color{fgcolor}
\includegraphics[width=\maxwidth]{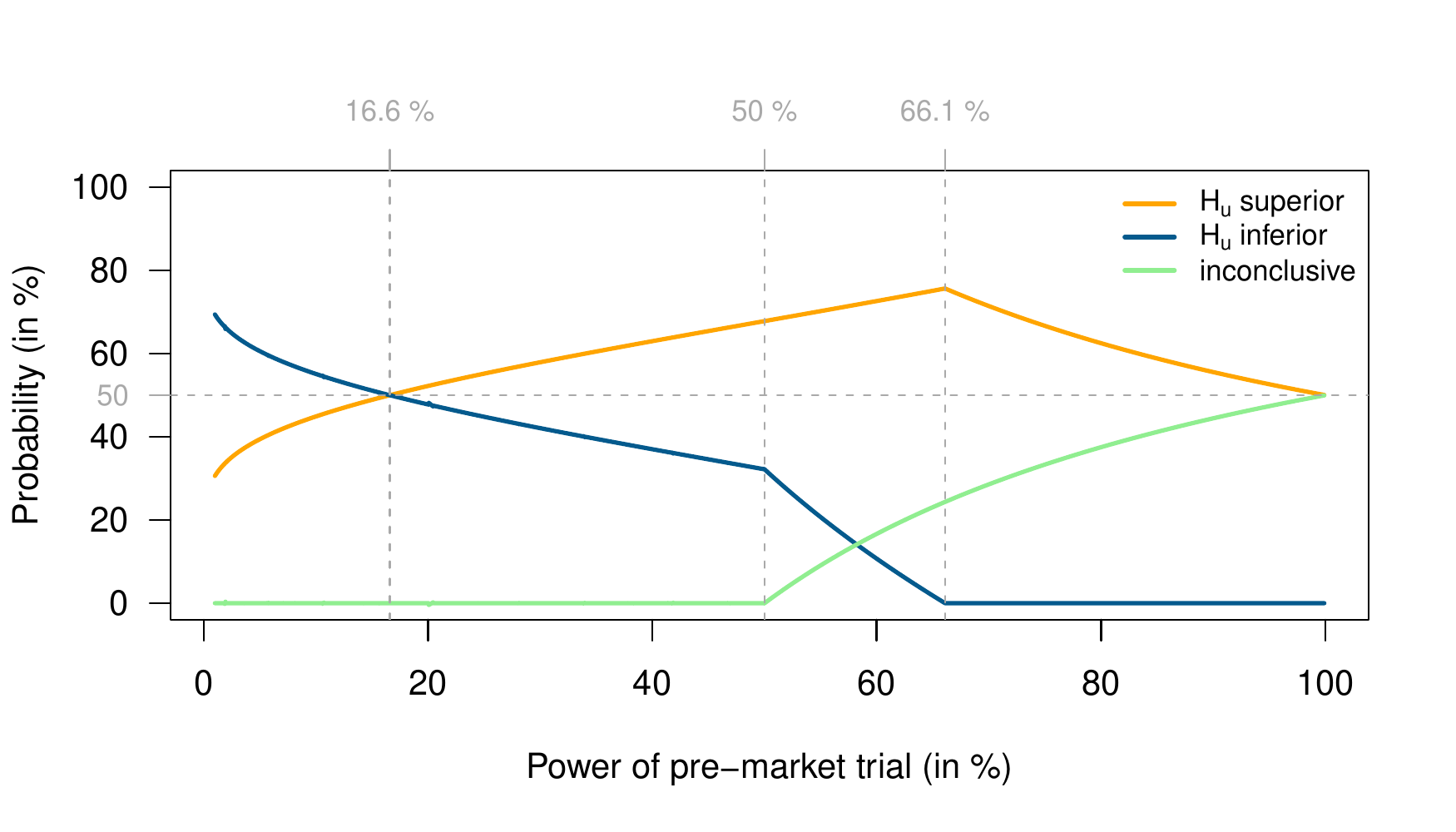} 
\end{knitrout}
\caption{Probability of the harmonic mean $\chi^2$-test to be superior/inferior to the two-trials rule
as a function of the true power from the pre-market clinical trial.}
\label{fig:methBetter}
\end{figure}

The results are depicted in Figure~\ref{fig:methBetter}.
For $\alpha = 0.025$,
the condition $\mu < b$ in~\eqref{eq:2trbetter} is fulfilled if
$1 - \beta < 66.1$\%  , so if the power
of the pre-market clinical trial is larger than
$66.1$\%, the harmonic mean $\chi^2$-test
is never inferior to the two-trials rule. Moreover, if the power
of the pre-market trial is 50\% or less, $z_1 > \mu$ always holds
(as $z_1 > z_{1 - \alpha}$ by construction), and so the probability of an
inconclusive result is null.
The probability that the harmonic mean $\chi^2$-test is inferior
to the the two-trials rule is larger
than 50\% only if the
pre-market clinical trial was severely underpowered to detect the true effect
$\theta$ (power of pre-market trial $< 16.6$\%). 
An extended Figure with a differentiation between the two types of inconclusive 
results can be found in the appendix (Figure~\ref{fig:methBetter_app}). 
Moreover, similar results can be obtained for the weighted harmonic mean $\chi^2$-test.

\section{POWER AT INTERIM \hlrev{OF POST-MARKET TRIAL}}
\label{PowerAtInterim}
The methods discussed above combine the final results of the 
pre- and post-market clinical trials. 
Another scenario is that the post-market clinical trial is still ongoing but the 
results of an interim analysis are already available. In this scenario, the 
power at interim can be calculated. The power at interim is ``the power of a 
[post-market clinical trial] taking into account the data from an 
interim analysis''.\citep{MicheloudHeld2022}$^{\text{(p.~370)}}$ If the power at 
interim is very low, 
it might be advisable to consider a premature termination of the clinical trial 
for futility.
The interim power can be the power to detect the effect $\hat\theta_1$
from the 
pre-market trial (conditional power, CPi), the power to detect 
$\hat\theta_1$
acknowledging its uncertainty (informed predictive power, IPPi) or can 
ignore the original result (predictive power, PPi).  
The description and formulas of the three types of interim power can be found \hlrev{in}
Micheloud and Held.\cite{MicheloudHeld2022}
We will focus 
on the IPPi in this work; we have adapted it
to the drug market licensing framework and applied it to the simulation 
study for illustrative purposes.

Two methods 
are used to calculate the interim power: the harmonic mean $\chi^2$-test
(we restrict on the unweighted case) and the two-trials rule. 
The interim power is thus either the probability that $p_2 \leq \bar p_2$ 
as calculated in \eqref{eq:threshold}
 (harmonic mean $\chi^2$-test) or the probability 
that $p_2 \leq \alpha$ (two-trials rule)
at the final analysis, given the samples collected so far at the time 
of the interim analysis.
We assume that the interim analysis occurs after half of the data was available; 
and so the sample size (per group) at interim $n_{2, i} = n_2/2$, with $n_2$ 
as in Figure~\ref{FigSampleSize}.
The test statistics $z_{2,i}$ at interim of the post-market clinical trial was 
simulated based on a normal distribution:

\begin{align*} 
z_{2,i} \sim \Nor\bigg(\theta_{2} \cdot \frac{\sqrt{n_{2,i}}}{\sqrt{2} 
\sigma}, 1\bigg).
\end{align*}

\begin{figure}[!h]
\centering
\begin{knitrout}
\definecolor{shadecolor}{rgb}{0.969, 0.969, 0.969}\color{fgcolor}
\includegraphics[width=\maxwidth]{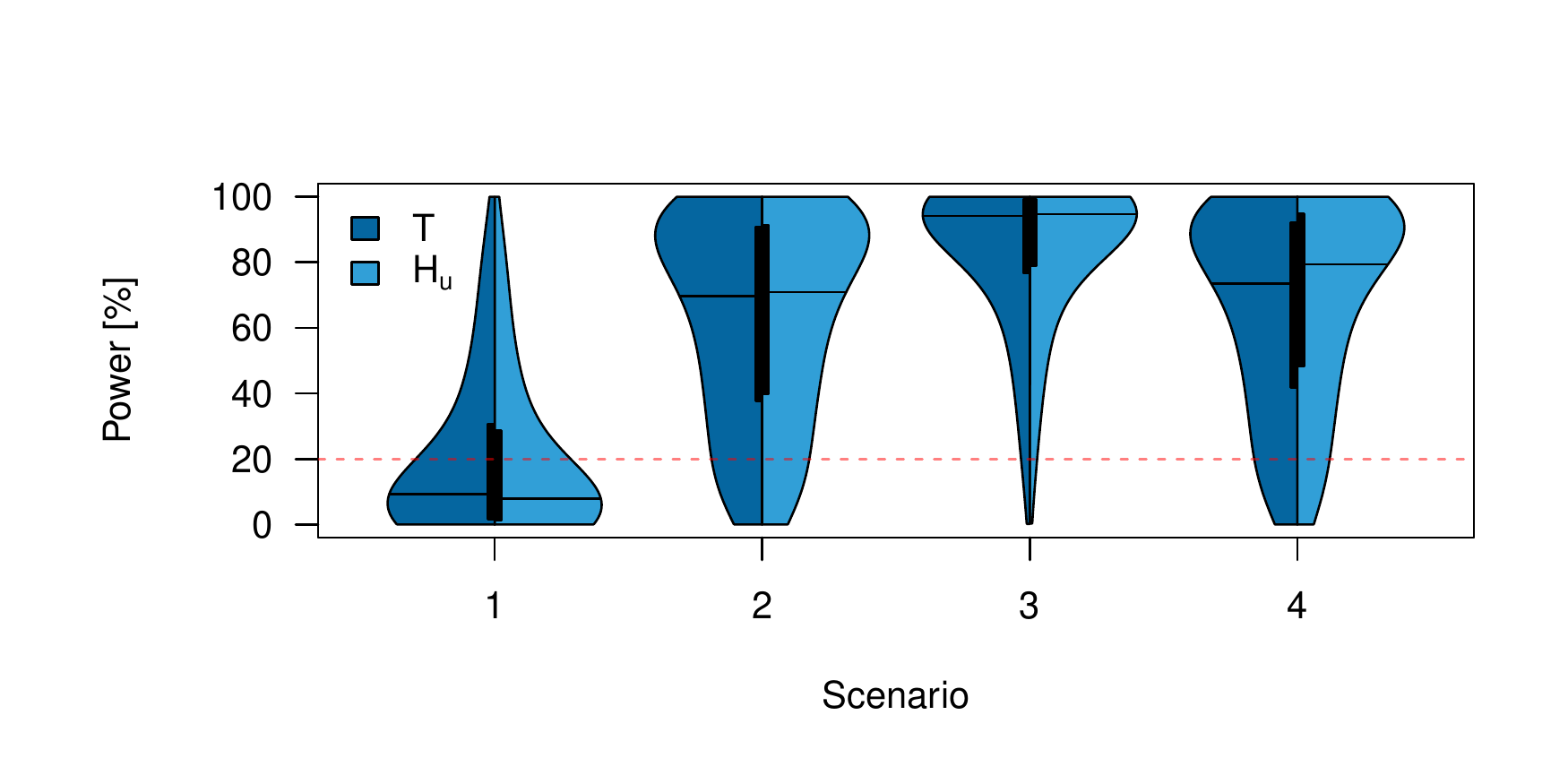} 
\end{knitrout}
\caption{Informed predictive predictive power with the 
two-trials rule ($T$) and the unweighted harmonic 
mean $\chi^2$-test ($H_u$) under the four scenarios.}
\label{fig:interim-power}
\end{figure}

The interim power under each scenario and with 
$n_{2, i} = n_2/2$ is presented in Figure~\ref{fig:interim-power}.
Some authors \cite{demets2006, bauer2006, MicheloudHeld2022}
suggest to use a futility
boundary of 20\%, \ie the trial should be stopped at interim 
if the interim power falls below 20\%. 
With this threshold, the proportion of clinical trials stopped for 
futility under the null (scenario 1) is 
67.7\% ($H_u$) and 
65.3\% ($T$), 
so more clinical trials would be correctly stopped using
the harmonic mean $\chi^2$-test rather than 
the two-trials rule. 
In contrast, the proportion of clinical trials (incorrectly) stopped 
at interim under the alternative (scenarios 2, 3 and 4) is smaller
with the harmonic mean $\chi^2$-test rather than 
with the two-trials rule (scenario 2: 
12.4\% ($H_u$) vs 
13.5\% ($T$), scenario 3: 
1.8\% ($H_u$) vs 
2.4\% ($T$) and 
scenario 4: 
9.2\% ($H_u$) vs 
11.5\% ($T$)). 
In addition, the median interim power is smaller with $H_u$ as compared 
to $T$ under the null, and larger under the alternative.  

\section{CONCLUSIONS}
\label{chap:Discussion}
This paper provides insights how to assess the combined evidence from clinical trials in terms of 
significance and sample size, which is relevant in settings for which 
conditional \hlrev{or accelerated} approval is considered appropriate. We evaluated the applicability of different statistical methods and, in particular,
we proposed the use of the harmonic mean $\chi^2$-test. In contrast to
Fisher's criterion or Stouffer's method, the proposed method 
always requires
a post-market trial to be conducted. Moreover, it takes into account the evidence from the
pre-market clinical trial by generating an adaptive significance level for the
post-market trial. In contrast to the two-trials rule, 
the harmonic mean $\chi^2$-test 
requires a smaller sample size for the post-market
clinical trial if the $p$-value $p_1$ from the pre-market clinical trial
is $\ll 0.025$ and a larger sample size if $p_1 \approx 0.025$ (for 
$\alpha = 0.025$). The proposed method
can also be used to calculate the power at interim of the
post-market clinical trial and can easily be extended to more than two
clinical trials.  When used for the design and the analysis of the 
post-market trial the proposed method tends to
be superior to the two-trials rule, \ie tends to
require a smaller sample size while still achieving larger power. 
\hlrev{These results allow, in addition to the independent analyses and 
assessments of the trials, a rational and quantitative assessment of the 
level of evidence that is consistent with the \hlrev{full approval} process.}

\hlrev{Recent work has shown that both the two-trials rule and the harmonic
mean $\chi^2$-test are special cases of a general family of
combination tests for two studies, indexed by the necessary bound
$\tilde \alpha$ for the study-specific $p$-values $p_1$ and $p_2$.
\citep{held_etal2022} The two-trials rule is represented by the
choice $\tilde \alpha = \alpha = 0.025$, where the necessary condition $p_1
\leq \tilde \alpha$ and $p_2 \leq \tilde \alpha$ is also sufficient. The
harmonic mean $\chi^2$-test corresponds to the necessary (but not
sufficient) bound $\tilde \alpha = 0.065$, see Section~\ref{MethodHarmonicMean}.
The general
framework allows for necessary bounds other than 0.065 for the
study-specific $p$-values $p_1$ and $p_2$. For example, we might want
to impose a more stringent necessary bound such as 0.05 or 0.04, but
not as strict as the two-trials rule where $\tilde \alpha = 0.025$.
Further research is needed to investigate the applicability of the
methods described in this paper to this more flexible setting. In
particular, the general framework \citep{held_etal2022} does currently not
allow for weights.} \hlrev{Future work could also focus on adapting this framework to cases where 
asymptotic normality does not hold, such as rare diseases where 
the sample sizes are small.}

\hlrev{In practice the role of replication by conducting 
a second study is not necessary to reach a combined error rate of $\alpha^2$ 
across studies. \hlrev{It is primarily to address the residual uncertainty on 
clinically relevant efficacy (and safety) that exists at conditional or 
accelerated approval.} The suggestion of assessing an overall error rate 
across trials 
also does not appear in guidance from regulators. 
The independent replication, preferably under different circumstances, 
is part of the 
scientific method as a whole \hlrev{and supports} confirming the robustness 
of observed effects. 
 For final regulatory assessment, 
achieving statistical significance under appropriate Type-I error control is in 
essence a ticket to entry of full approval,  because there is a clear signal 
of confirmed efficacy. Full assessment of all aspects (efficacy and safety in the
broad sense) ultimately leads to regulatory decisions for approval, which are rarely 
simple acceptance or not: in almost all cases refinements  in indication and target 
population(s) are included, as well as conditions for manufacturing and use are part of
the decision.}

%

\section{Appendix}

\subsection{\hl{Detailed version of  Figure~\ref{fig:methBetter}}}
\begin{figure}[!h]
\centering
\begin{knitrout}
\definecolor{shadecolor}{rgb}{0.969, 0.969, 0.969}\color{fgcolor}
\includegraphics[width=\maxwidth]{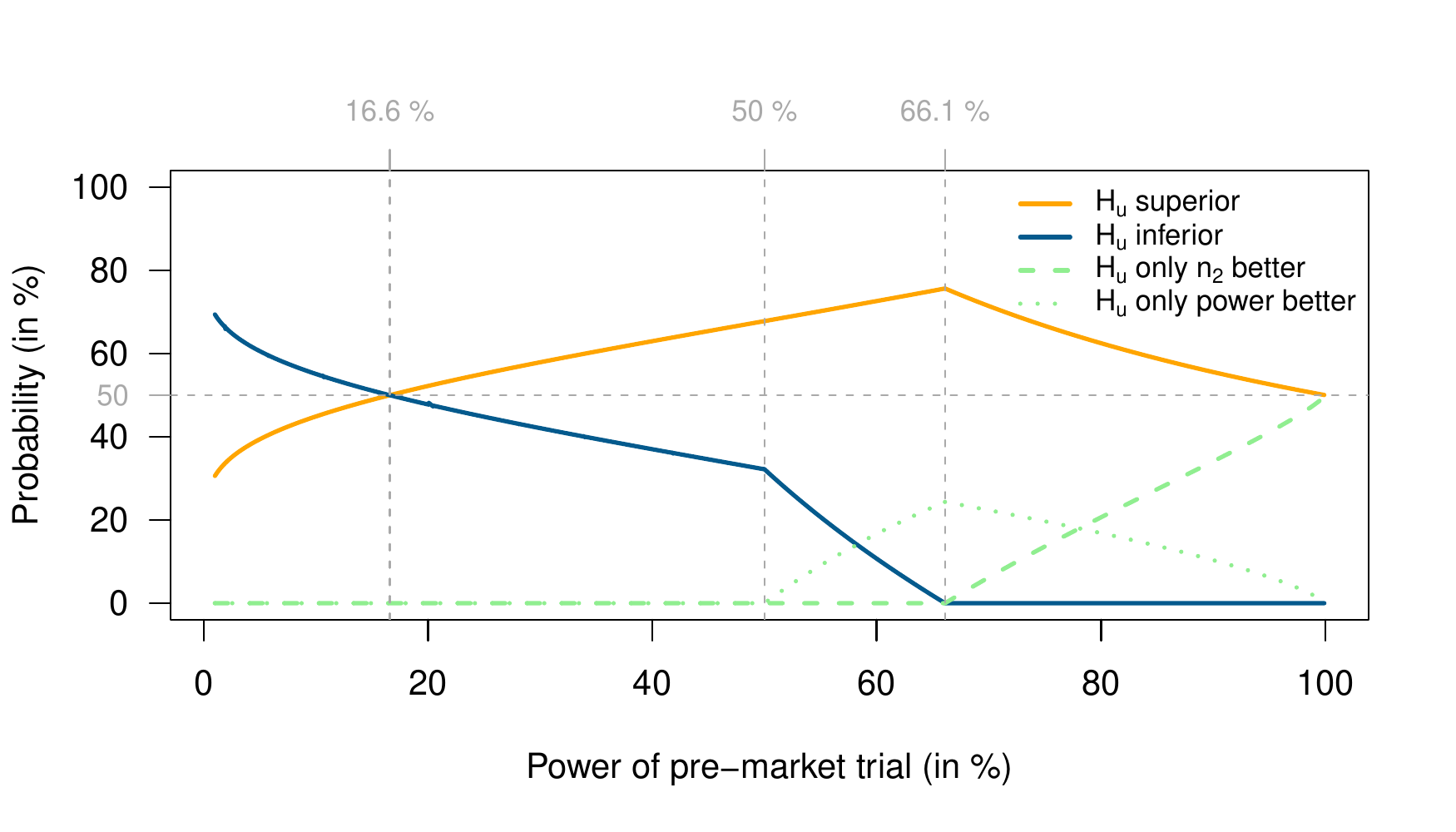} 
\end{knitrout}
\caption{Probability of the harmonic mean $\chi^2$-test to be superior/inferior 
to the two-trials rule as a function of the true power from the pre-market 
clinical trial.}
\label{fig:methBetter_app}
\end{figure}

Figure~\ref{fig:methBetter_app} is a more \hl{detailed} version 
of Figure~\ref{fig:methBetter}, where the inconclusive results are 
separated into two categories: the harmonic mean $\chi^2$-test method is 
superior to the two-trials rule either with respect to the sample size $n_2$ 
(smaller) or the power (larger). 


\nocite{*}
\bibliographystyle{ama}
\bibliography{bibliography}

\begin{thebibliography}{10}

\bibitem{EMA2016Guideline}
{European Medicines Agency} {(EMA)}. Guideline on the scientific application
  and the practical arrangements necessary to implement Commission Regulation
  (EC) No 507/2006 on the conditional marketing authorisation for medicinal
  products for human use falling within the scope of Regulation (EC) No
  726/2004
  \url{https://www.ema.europa.eu/en/documents/scientific-guideline/guideline-scientific-application-practical-arrangements-necessary-implement-commission-regulation-ec/2006-conditional-marketing-authorisation-medicinal-products-human-use-falling_en.pdf}
  2016.
\newblock Accessed 05 Oct 2021. Last update 25 Feb 2016.

\bibitem{FDAGuidance}
{FDA} . Demonstrating substantial evidence of effectiveness for human drug and
  biological products: Guidance for Industry
  \url{https://www.fda.gov/media/133660/download} 2019.
\newblock Draft guidance. Accessed 19 Feb 2020. Last update December 2019.

\bibitem{EMAFampyra}
{European Medicines Agency} {(EMA)}. Fampyra
  \url{https://www.ema.europa.eu/en/medicines/human/EPAR/fampyra#overview-section}
  2020.
\newblock Accessed 13 Sep 2021. Last update 24 Aug 2020.

\bibitem{Fampyra:AssessReport}
{European Medicines Agency} {(EMA)}. Assessment report {F}ampyra
  \url{https://www.ema.europa.eu/en/documents/assessment-report/fampyra-epar-public-assessment-report_en.pdf}
  2011.
\newblock First published 04 Aug 2011. Last update 23 Jun 2011.

\bibitem{Hobart2019}
Hobart J, Ziemssen T, Feys P, et al. Assessment of clinically meaningful
  improvements in self-reported walking ability in participants with Multiple
  Sclerosis: Results from the randomized, double-blind, Phase III ENHANCE trial
  of prolonged-release Fampridine  {\it CNS Drugs. } 2019;33:61-79.

\bibitem{Altman.etal2000}
Altman D~G, Machin D, Bryant T~N, Gardner M~J. {\it {Statistics with
  Confidence}}.
\newblock BMJ Books~2nd~ed. 2000.

\bibitem{Held2019}
Held L. The harmonic mean $\chi^2$-test to substantiate scientific findings
  {\it Journal of the Royal Statistical Society: Series C (Applied Statistics).
  } 2020;69:697-708.

\bibitem{Spiegelhalter2004}
Spiegelhalter D~J, Abrams K~R, Myles J~P. {\it Bayesian Approaches to Clinical
  Trials and Health-Care Evaluation}.
\newblock Chichester: John Wiley \& Sons Ltd 2004.

\bibitem{FDA1998}
{U.S. Food and Drug Administration} {(FDA)}. Providing clinical evidence of
  effectiveness for human drug and biological products.   1998.

\bibitem{PtC}
Proprietary Medicinal Products~{CPMP} {EMA}. Points to consider on application
  with 1. Meta-analyses; 2. One pivotal study
  \url{https://www.ema.europa.eu/en/application-1-meta-analyses-2-one-pivotal-study}
  2001.

\bibitem{ioannidis2008}
Ioannidis J~P~A. Why most discovered true associations are inflated  {\it
  Epidemiology. } 2008;19:640--648.

\bibitem{Rothwell2021}
Rothwell J~C, Julious S~A, Cooper C~L. Adjusting for bias in the mean for
  primary and secondary outcomes when trials are in sequence  {\it
  Pharmaceutical Statistics. } 2021;21:460--475.

\bibitem{MicheloudHeld2022}
Micheloud C, Held L. {Power calculations for replication studies}  {\it
  Statistical Science. } 2022;37:369 -- 379.

\bibitem{Matthews2006}
Matthews J~N~S. {\it {I}ntroduction to {R}andomized {C}ontrolled {C}linical
  {T}rials}.
\newblock Chapman \& Hall/CRC~2nd~ed. 2006.

\bibitem{Fisher-1958}
Fisher R~A. {\it {Statistical Methods for Research Workers}}.
\newblock Edinburgh: Oliver \& Boyd~13th ed. (rev.)~ed. 1958.

\bibitem{Bauer1994}
Bauer P, Köhne K. Evaluation of experiments with adaptive interim analyses
  {\it Biometrics. } 1994;50:1029-1041.

\bibitem{Fisher1999}
Fisher L~D. One large, well-designed, multicenter study as an alternative to
  the usual FDA paradigm  {\it Drug Information Journal. } 1999;33:265--271.
\newblock \url{https://doi.org/10.1177/009286159903300130}.

\bibitem{Shun2005}
Shun Z, Chi E, Durrleman S, Fisher L. Statistical consideration of the strategy
  for demonstrating clinical evidence of effectiveness{\textemdash}one larger
  vs two smaller pivotal studies  {\it Statistics in Medicine. }
  2005;24:1619--1637.

\bibitem{Burton2006}
Burton A, Altman D~G, Royston P, Holder R~L. The design of simulation studies
  in medical statistics  {\it Statistics in Medicine. } 2006;25:4279-4292.

\bibitem{Morris2019}
Morris T~P, White I~R, Crowther M~J. Using simulation studies to evaluate
  statistical methods  {\it Statistics in Medicine. } 2019;38:2074-2102.

\bibitem{R}
R~{Core Team}. {\it R: A Language and Environment for Statistical Computing}.
\newblock R Foundation for Statistical Computing~Vienna, Austria 2021.

\bibitem{button2013}
Button K~S, Ioannidis J~P~A, Mokrysz C, et al. Power failure: why small sample
  size undermines the reliability of neuroscience  {\it Nature Reviews
  Neuroscience. } 2013;14:365.

\bibitem{Tanniou2016}
Tanniou J, {van der Tweel} I, Teerenstra S, Roes K~C~B. Level of evidence for
  promising subgroup findings in an overall non-significant trial  {\it
  Statistical Methods in Medical Research. } 2016;25:2193--2213.

\bibitem{demets2006}
DeMets D~L. Futility approaches to interim monitoring by data monitoring
  committees  {\it Clinical Trials. } 2006;3:522--529.

\bibitem{bauer2006}
Bauer P, K{\"o}nig F. The reassessment of trial perspectives from interim
  data--a critical view  {\it Statistics in Medicine. } 2006;25:23--36.
\newblock \url{https://doi.org/10.1002/sim.2180}.

\bibitem{held_etal2022}
Held L, Micheloud C, Balabdaoui F. A statistical framework for replicability.
  tech. rep. 2022.
\newblock \url{https://arxiv.org/abs/2207.00464}.

\bibitem{EMA2021}
{European Medicines Agency} {(EMA)}. EMA receives application for conditional
  marketing authorisation of {COVID}-19 {V}accine {A}stra{Z}eneca
  \url{https://www.ema.europa.eu/en/news/ema-receives-application-conditional-marketing-authorisation-covid-19-vaccine-astrazeneca}
  2021.
\newblock Accessed 20 Sep 2021. Last update 12 Jan 2021.

\end{thebibliography}
\end{document}